\documentclass[runningheads]{llncs}
\usepackage{graphicx}
\usepackage{amsfonts}

\usepackage{xcolor}
\usepackage[detect-weight=true,detect-family=true]{siunitx}
\usepackage{xspace}

\usepackage{amsmath}

\usepackage{tikz}
\usetikzlibrary{positioning}

\usepackage{multirow}
\usepackage{hyperref}
\usepackage{textgreek}

\definecolor{darkgrey}{RGB}{80,80,80}
\definecolor{lightgrey}{RGB}{170,170,170}

\definecolor{brown}{HTML}{a52a2a}
\definecolor{darkcyan}{HTML}{0a888a}

\usepackage{adjustbox}
\usepackage{booktabs}
\usepackage{wasysym}
\usepackage{tabularx} 
\usepackage{multirow}
\usepackage{arydshln} 
\usepackage{siunitx}

\newcommand\Tstrut{\rule[-5pt]{0pt}{14pt}}         

\newcommand{\ie}{\textit{i.e.},\xspace}
\newcommand{\eg}{\textit{e.g.},\xspace}
\newcommand{\cf}{\textit{cf.}\xspace}

\newcommand{\wrt}{w.r.t.\xspace}

\newcommand{\code}[1]{\texttt{#1}}

\newcommand{\CUT}[1]{{\iffalse#1\fi}}

\usepackage[nolist,nohyperlinks]{acronym}
\begin{acronym}[TDMA]
	\acro{ICS}{Industrial Control System}
	\acro{MitM}{Man-in-the-Middle}
	\acro{WSN}{Wireless Sensor Network}
	\acro{PER}{Packet Error Rate}
	\acro{CPS}{Cyber-physical System}
	\acro{DoS}{Denial of Service}
	\acro{IoT}{Internet of Things}
	\acro{MAC}{Message Authentication Code}
\end{acronym}

\usepackage{layouts}
\usepackage{xfrac}
\usepackage{textpos}

\newcommand{\agg}{\texttt{Agg$(\cdot)$}\xspace}
\newcommand{\comp}{\texttt{Comp$(\cdot)$}\xspace}
\newcommand{\sw}{\texttt{SW$(\cdot)$}\xspace}
\newcommand{\rd}{\texttt{R2D2$(\cdot)$}\xspace}

\begin{document}

\title{When and How to Aggregate Message Authentication Codes on Lossy Channels?}
%
%
\author{
	Eric Wagner\inst{1,2} \and
	Martin Serror\inst{1} \and
	Klaus Wehrle\inst{2} \and
	Martin Henze\inst{3,1}
}
\authorrunning{E.\ Wagner et al.}
%
\institute{
	Cyber Analysis \& Defense, Fraunhofer FKIE, \\ Wachtberg, Germany \email{\{firstname.lastname\}@fkie.fraunhofer.de} \and
	Communication and Distributed Systems, RWTH Aachen University, \\ Aachen, Germany \email{\{lastname\}@comsys.rwth-aachen.de} \and
	Security and Privacy in Industrial Cooperation, RWTH Aachen University, \\ Aachen, Germany
\email{henze@spice.rwth-aachen.de}
}

\titlerunning{When and How to Aggregate Message Authentication Codes?}

%
%
\maketitle              

\begin{abstract}

Aggregation of message authentication codes (MACs) is a proven and efficient method to preserve valuable bandwidth in resource-constrained environments: 
Instead of appending a long authentication tag to each message, the integrity protection of multiple messages is aggregated into a single tag.
However, while such aggregation saves bandwidth, a single lost message typically means that authentication information for multiple messages cannot be verified anymore.
With the significant increase of bandwidth-constrained lossy communication, as applications shift towards wireless channels, it thus becomes paramount to study the impact of packet loss on the diverse MAC aggregation schemes proposed over the past 15 years to assess when and how to aggregate message authentication.
Therefore, we empirically study all relevant MAC aggregation schemes in the context of lossy channels, investigating achievable goodput improvements, the resulting verification delays, processing overhead, and resilience to denial-of-service attacks.
Our analysis shows the importance of carefully choosing and configuring MAC aggregation, as selecting and correctly parameterizing the right scheme can, e.g., improve goodput by \SI{39}{\percent} to \SI{444}{\percent}, depending on the scenario.
However, since no aggregation scheme performs best in all scenarios, we provide guidelines for network operators to select optimal schemes and parameterizations suiting specific network settings.


\end{abstract}

\section{Introduction}
\label{sec:intro}

With the proliferation of the (industrial) \ac{IoT}, more and more battery-operated devices, such as sensors and actuators, rely on wireless communications.
Consequently, the number of devices sharing the same transmission medium (with a fixed capacity) is growing, imposing increasingly stringent bandwidth constraints on \ac{IoT} applications~\cite{2019_vitturi_industrial}.
At the same time, wireless communication further amplifies the need to adequately secure transmitted messages~\cite{2021_serror_challenges}, most notably to ensure the integrity of transmitted critical information~\cite{2017_castellanos_retrofitting2}, which would have prevented \eg the 2015 and 2016 cyberattacks on the Ukrainian power grid~\cite{2017_whitehead_ukraine}.
However, establishing integrity protection requires additional bandwidth to transmit authentication tags, thus conflicting with the already hard-to-reach constraints of \ac{IoT} communication.
Therefore, a vital research topic for industry and academia centers around the question of how to use the shared limited transmission resources efficiently and still provide adequate security~\cite{2020_seferagic_survey}.

As a result, many efforts across protocol stacks have been proposed to reduce bandwidth overhead. 
Prominent examples include 6LoWPAN header compression~\cite{2012_raza_6lowpan} or, more recently, the record layer headers of DTLS 1.3~\cite{RFC9147} and Compact TLS 1.3~\cite{ctls_draft}.
Such protocol improvements can however not address the inherent overhead necessary to provide integrity protection.
Considering, \eg desirable 128-bit security requires the integration of a 16-byte authentication tag into the message's payload.
Moreover, since (industrial) \ac{IoT} protocols such as IEEE 802.15.4, LoRaWAN, or Bluetooth Low Energy often rely on short messages, such \acp{MAC} typically occupy a significant portion of each message and, in some cases, do not even fit~\cite{2008_nilsson_compoundmac}.

For at least 15 years, the well-established and time-proven concept of MAC aggregation has been known to alleviate these limitations~\cite{2008_katz_aggregated-mac}.
The idea is simple yet effective: 
Instead of protecting the integrity of each message individually, a single authentication tag is responsible for protecting the integrity of multiple messages.
Given a reliable channel, this approach works flawlessly and can be reduced to a trade-off between saved bandwidth and the verification delay for received messages:
Aggregating integrity protection of more and more messages reduces the induced overhead until it becomes negligible but implies that the receiver has to wait for the reception of all messages affected by the aggregation before being able to check their integrity, resulting in significant delays if too many authentication tags are aggregated.

Over the years, different MAC aggregation schemes have been proposed to address weaknesses~\cite{2010_eikemeier_history,2012_kolesnikov_multiplicity,2018_hirose_non}, split authentication tags over multiple messages~\cite{2008_nilsson_compoundmac}, or provide progressive security guarantees~\cite{2020_armknecht_promac,2021_li_cumac,2022_wagner_spmac}.
And while various implementations of security concepts, such as message authentication~\cite{2021_simplicio_survey}, have been evaluated and compared by literature, such analyses of \ac{MAC} aggregation schemes in realistic wireless, and thus lossy, settings are practically non-existent.
Most importantly, current evaluations of MAC aggregation schemes neglect that losing a single message from a set of messages with aggregated authentication tags may have cascading effects depending on the chosen \ac{MAC} aggregation scheme.
This phenomenon becomes increasingly relevant as more and more communication transitions to low-bandwidth wireless, and thus lossy, channels in a diverse set of applications such as smart cities, underwater communication, or the (industrial) IoT~\cite{2016_chen_performance}.
Thus, \ac{MAC} aggregation is arguably becoming even more critical for lossy channels than for its initial setting of reliable communication.
However, research, thus far, did not provide sufficient address under which circumstances \ac{MAC} aggregation on lossy channels is sensible and how to unlock its full potential.
This knowledge is, however, crucial to optimally utilize scarce bandwidth in wireless scenarios with an ever-growing number of participating devices.

To address these shortcomings, this paper addresses the hitherto neglected analysis of the performance of relevant MAC aggregation schemes in the presence of lossy channels. 
We consider realistic wireless (industrial) \ac{IoT} communication scenarios, which suffer from scarce transmission resources and significant packet losses, where we compare the performance of existing MAC aggregation schemes. 
Our analysis is thus a valuable contribution for security practitioners and researchers: On the one hand, it allows identifying suitable aggregation schemes depending on the considered scenario, and on the other hand, it reveals current shortcomings, which lay the foundation for identifying more effective approaches.
Ultimately, we want to answer the questions of when \ac{MAC} aggregation is sensible on lossy channels and how this aggregation should be performed by making the following contributions:

\begin{itemize}
	\item We investigate the achievable goodput improvements of all MAC aggregation scheme known to us under various parameterizations in synthetic and real-world scenarios~(Section~\ref{sec:synthetic} and Section~\ref{sec:real});
	\item We further analyze the impact of MAC aggregation on decisive factors such as verification delay, processing times, memory cost, and the susceptibility to denial-of-service attacks~(Section~\ref{sec:beyond}); and
	\item Finally, we provide actionable guidelines to help in deciding when and how current MAC aggregation schemes are best deployed~(Section~\ref{sec:guidelines}).
\end{itemize}

\textbf{Availability Statement.} To help in the decision process of which, if any, MAC aggregation scheme should be deployed in a concrete scenario, our tool to compare MAC aggregations schemes in concrete scenarios is available at: \url{https://github.com/fkie-cad/mac-aggregation-analysis-tool}

\section{MAC Aggregation on Lossy Channels}
\label{{sec:background}}

Achieving integrity protection is a significant challenge in bandwidth-constrained environments.
Even the tiniest message requires an authentication tag of several bytes (e.g., 16 bytes for 128-bit security), thus occupying considerable space in each message.
\ac{MAC} aggregation schemes, as presented in this section, try to alleviate this overhead by distributing the burden of authentication over multiple messages.
In the following, we first define \acp{MAC}~(Section~\ref{sec:background:mac}) before formally introducing the concept of \ac{MAC} aggregation~(Section~\ref{sec:background:formalizing}).
To conclude, we introduce the existing \ac{MAC} aggregation schemes~(Section~\ref{sec:background:schemes}) and motivate research into their applicability in lossy conditions~(Section~\ref{sec:background:motivation}).

\subsection{Message Authentication Codes}
\label{sec:background:mac}

\acfp{MAC} allow two communication partners to verify the integrity of exchanged messages using a pre-shared secret $k$~\cite{2023_boneh_crypto}.
This key $k$ can be derived dynamically through a key exchange protocol or hardcoded at both communicating entities.
To authenticate a message $m$, the sender uses the tag generation algorithm $\textit{Sig}_k(m)$ to generate the corresponding authentication tag~$t$.
Upon reception of a message, the verification algorithm $\textit{Vrfy}_k(m,t)$ enables the recipient to evaluate whether the received tag is valid. 
Typically, this verification is done by computing the tag $t^* = {Sig}_k(m^*)$ for the received message $m^*$ and comparing it to the received tag $t$.
A \ac{MAC} scheme is considered secure if it is computationally infeasible to generate a ($m$,$t$)-pair that $\textit{Vrfy}_k(\cdot)$ would accept without knowing the secret $k$.
This requirement can be achieved by, \eg using keyed hash functions such as \code{HMAC-SHA256} to compute $t$.
Thus, \acp{MAC} provide integrity protection for communication channels, where they prevent any attacker not knowing $k$ from undetectably manipulating the content of transmitted messages.

\subsection{MAC Aggregation to Combat Bandwidth Scarcity}
\label{sec:background:formalizing}

Traditional \ac{MAC} schemes consume significant bandwidth in constrained environments.
Over 15 years ago, the concept of \ac{MAC} aggregation was promoted to combat these limitations~\cite{2008_katz_aggregated-mac}.
The idea is elegant and effective: Instead of authenticating each message individually, a single tag is responsible for protecting the integrity of multiple messages.
Thus, the overhead of each tag is distributed over multiple messages, saving valuable bandwidth.

Formally, \ac{MAC} aggregation schemes can be defined as an extension of traditional \ac{MAC} schemes.
In a traditional \ac{MAC} scheme, the tag $t_i$ is computed over and transmitted alongside message $m_i$.
For \ac{MAC} aggregation schemes, the aggregated tag $t_i^{\text{agg}}$, which is transmitted alongside $m_i$, is computed by aggregating the integrity protection of multiple messages $m_{i-d} (d\in\mathcal{D})$ with an additional keyless function $\textit{Agg}(\cdot)$, such that $t_i^{\text{agg}} = \textit{Agg}(t_{i-d} | d\in\mathcal{D})$.
We say that $\mathcal{D}\subset\mathbb{N}_0$ is the set of dependencies of a \ac{MAC} aggregation scheme and, \eg $2\in\mathcal{D}$ means that the tag $t_{i-2}$ is included in the computation of the aggregated tag $t_i^{\text{agg}}$.
Vice versa, the integrity of message $m_{i-2}$ is protected by tags $t_{i}$.

Thus, aggregated authentication tags protect multiple messages.
At the same time, each message is potentially protected by multiple tags as each (potentially shortened) tag may only be responsible for providing a fraction of the overall targeted security level.
Since each tag aggregates integrity protection for multiple messages, aggregated \ac{MAC} schemes result in, on average, shorter tags.
In this context, the dependencies $\mathcal{D}$ describe how the reception of one message influences the verifiability of tags and the authenticity of surrounding messages.
We say that if an aggregated \ac{MAC} scheme has the dependencies $\mathcal{D}$, the generation and verification of tag $t_i$ require knowledge of $\{ m_{i-d} | d\in\mathcal{D} \}$, as $t_{i-d} = \textit{Sig}_k( m_{i-d} )$.
Consequently, a message $m_i$ blends into all tags $\{ t_{i+d} | d\in\mathcal{D} \}$, and a tag $t_i$ protects the integrity of all messages $\{ m_{i-d} | d\in\mathcal{D} \}$.

A specific \ac{MAC} aggregation scheme defines the underlying \ac{MAC} scheme, the dependencies $\mathcal{D}$, and the aggregation function $\textit{Agg}(\cdot)$.
In the following, we consider a simple XOR of authentication tags for the aggregation function, \ie $t_i^{\text{agg}} = \textit{Agg}(t_{i-d} | d\in\mathcal{D}) = \bigoplus_{d \in \mathcal{D}} t_{i-d}$.
This aggregation of tags is efficient and has been shown to be secure~\cite{1995_bellare_xormac}.
If, for example, $t_i$ and $t_j$ provide 128-bit integrity protection for $m_i$ and $m_j$, then $t^\text{agg}=t_i\oplus t_j$ provides 128-bit integrity protection for both messages $m_i$ and $m_j$.
However, the security of this aggregation function requires that the chosen \ac{MAC} function is pseudorandom and includes a nonce for replay protection to prevent mix-and-match attacks within one set of jointly authenticated messages~\cite{2010_eikemeier_history}.
Consequently, \ac{MAC} schemes based on universal hashing, such as UMAC~\cite{1999_black_umac}, should not be used in combination with XOR-based \ac{MAC} aggregation\footnote{BP-MAC~\cite{2022_wagner_bpmac} (based on a Carter-Wegman construction), for example, is insecure if used in combination with XOR-based \ac{MAC} aggregation. As each bit is authenticated individually and replay protection is only provided through a blinding tag, an attack can undetectably swap the x-th bits' values of two messages.}.
Most prominent \ac{MAC} schemes, such as \code{HMAC-SHA256}, can, however, be securely used with XOR-based \ac{MAC} aggregation if used in combination with nonce-based replay protection.

\subsection{Introducing Existing MAC Aggregation Schemes}
\label{sec:background:schemes}

After formalizing the concept of \ac{MAC} aggregation in Section~\ref{sec:background:formalizing}, we now introduce the different sets of MAC aggregation schemes, grouped by their choice of dependencies $\mathcal{D}$ and computation of $t^{\text{agg}}$.
We do, however, not focus on the exact aggregation function or the underlying MAC scheme, as those choices do not impact the scheme's susceptibility to packet loss.
Under these aspects, we present all four classes of aggregation that cover, to the best of our knowledge, all proposed schemes.
For this presentation, we assume XOR-based aggregation with \code{HMAC-SHA256} as a suitable MAC scheme (including an appended nonce for replay protection).

\subsubsection*{Traditional (Trad.):}
To quantify the performance of existing \ac{MAC} aggregation schemes, we compare them to the baseline performance of traditional \ac{MAC} schemes.
Therefore, we consider a traditional \ac{MAC} scheme that authenticates each message $m_i$ with an individual tag $t_i$.
This computation thus solely depends on $m_i$, \ie $\mathcal{D}=\{0\}$.
As we target 128-bit security, the {HMAC-SHA256} is truncated to \SI{16}{\byte}.

\subsubsection*{Aggregated MAC (Agg(n)):}
The most prominent scheme is aggregated MAC as introduced in 2008~\cite{2008_katz_aggregated-mac} and later extended to prevent reordering attacks~\cite{2010_eikemeier_history}, allow messages to occur multiple times~\cite{2012_kolesnikov_multiplicity}, and identify faulty messages in an aggregate~\cite{2018_hirose_non}.
For these schemes, a tag $t^{\text{agg}}$ is only appended to each n-th message, where $n$ is the parameter for how many messages' authentication tags are aggregated together.
For our evaluation, we consider the aggregation of two, four, eight, and sixteen tags, \ie $n\in{2,4,8,16}$ to cover a range of different parameterizations.
For every n-th message, a tag is then computed by XORing the authentication tags of all considered messages, as formalized in the following:

\begin{equation*}
	t_i^{\text{agg}} = \underset{i-n < k \leq i}{\bigoplus} t_{k} \quad \text{for } i\equiv 0\pmod{n}
\end{equation*}

\subsubsection*{Compound MAC (Comp(n)):}
As the tags computed by \agg are too long for some applications, Compound MAC is proposed that splits across multiple messages~\cite{2008_nilsson_compoundmac}.
Thus, each message carries a shortened authentication tag, the length of which is inversely proportional to the number of aggregated messages, \ie $|t| = \sfrac{128}{n}$.
For our analysis, we again consider $n\in{2,4,8,16}$.
We formalize \comp in the following, where $t[a:b]$ means the chunk from the a-th to the b-th bit of tag $t$: 
\begin{equation*}
	t_i^{\text{agg}} = \underset{ \lfloor\frac{i}{n}\rfloor\cdot (n-1) \leq k < \lfloor\frac{i}{n}\rfloor\cdot n}{\bigoplus} t_{k}[(k\bmod n)\cdot |t| :((k+1)\bmod n)\cdot |t|]
\end{equation*}

\subsubsection*{Sliding Window-based Progressive MACs (SW(n,o)):}
Progressive MAC has been introduced to provide initially reduced security that is improved eventually upon message reception~\cite{2020_armknecht_promac,2021_li_cumac,2017_schmandt_minimac}.
Therefore, each message is protected by a shortened tag that also verifies the integrity of the previous $n$ messages.
As \sw is not equipped to provide full security under packet loss, it can be compensated by additionally considering an overprovisioning factor $o$.
This factor defines in percent how much security may be extended beyond the target, \ie $o=100$ means that messages may have 256-security at the expense of longer tags as $ |t^{\text{agg}}| = \sfrac{128}{n} \cdot (1+\sfrac{o}{100}) $.
Here, we select a number of parameter combinations that perform best under various scenarios.
The tag computation of \sw can be formalized as follows:
\begin{equation*}
	t_i^{\text{agg}} = \underset{i-n < k \leq i}{\bigoplus} t_{k}[k\cdot|t|:(k+1)\cdot|t|]
\end{equation*}

\subsubsection*{Randomized and Resilient Dependency Distribution (R2D2(n,g,o)):}
To address weaknesses of \sw in the presence of packet loss, \rd introduces dependencies that bound the effect a dropped packet can have on the verifiability of any other message~\cite{2022_wagner_spmac}.
Therefore, the parameter $g$ is introduced, which defines how much security any message loses at most if a surrounding packet is lost, \ie $g=1$ in combination with \SI{2}{\byte} long packets means that any message can lose at most 16 bit of security.
Furthermore, \rd randomizes the concrete dependency set $\mathcal{D}$ and assigns a different set to each bit of a tag.
The final aggregate tag $t^{\text{agg}}$ is thus a juxtaposition of bit-long tags $t^{\text{agg}}_j$ and is defined as:

\begin{equation*}
t^{\text{agg}}_j = \bigoplus_{0\leq k<|\mathcal{D}_j|} t_{i-\mathcal{D}_j[k]}[k*|t|+i]
\end{equation*}
with $\mathcal{D}_j[n]$ representing the $n$-th entry of j-th bit dependency set $\mathcal{D}_j$.

\subsection{Interplay of Lossy Channels and MAC Aggregation}
\label{sec:background:motivation}

\ac{MAC} aggregation can bring benefits to a wide range of constrained environments, such as \acp{ICS}, smart homes, smart city, or underwater networks.
However, we see these targeted environments quickly shifting towards more and more lossy communication with protocols such as ZigBee, Sigfox, Bluetooth Low Energy, or LoRaWAN, to name only a few.
This shift can significantly impact the performance of \ac{MAC} aggregation schemes, especially considering \acp{PER} that can rise to \SI{10}{\%} and above for certain scenarios~\cite{2022_wagner_spmac}.
With \ac{MAC} aggregation, a lost packet means that the receiver cannot authenticate the initially transmitted message and all other messages that depend on it.
Arguably, \ac{MAC} aggregation has become even more critical in the lossy settings than for reliable communications since these networks more often expose bandwidth constraints due to the high number of nodes sharing the same transmission medium.
LoRaWAN, for example, is often limited to less than \SI{10}{\kilo\byte} or even \SI{1}{\kilo\byte} of throughput per hour per device.
Despite this stringent requirement of conserving bandwidth in lossy networks, no accurate performance analysis of \ac{MAC} aggregation in this context has been conducted  thus far to the best of our knowledge.
In the following section, we provide the first such analyses for the different \ac{MAC} aggregation schemes presented in Section~\ref{sec:background:schemes}.

\section{Synthetic Measurements}
\label{sec:synthetic}

We begin our analyses of \ac{MAC} aggregations schemes by looking at synthetic measurements of simulated wireless channels.
These measurements give us fine control over channel quality and payload length to investigate how these parameters influence the different \ac{MAC} aggregation schemes.
In Section~\ref{sec:synthetic:setup}, we first describe our setup before diving into the influence of channel quality and payload lengths in Sections~\ref{sec:synthetic:power}~and~\ref{sec:synthetic:length}, respectively.
Finally, we look at the established challenge of determining optimal payload lengths for given channel qualities under the additional constraint that the received data must be authenticated.

\subsection{Simulation Setup}
\label{sec:synthetic:setup}

For our synthetic measurements, we use the network simulator \texttt{ns-3} (version 3.37), giving us fine-grained control over the underlying communication channel.
As communication protocol, we choose the IEEE~802.15.4 protocol commonly used in constrained wireless environments and included in \texttt{ns-3}, where we consider the most compact header of \SI{5}{\byte}.
For payload lengths varying between \SI{1}{\byte} and the maximum supported \SI{115}{\byte}, we simulate the communication between two static antennas placed \SI{25}{\meter} apart and extract binary loss traces of which transmitted packets have been correctly received or not.
We additionally vary the transmit power varying between \SI{-21}{dBm} and \SI{-16}{dBm} using \SI{.1}{dBm} steps to increase the signal-to-noise ratio progressively, thus improving the channel quality and reducing \ac*{PER}.
We only transmit each message once and do implement acknowledgments or retransmission, as these features are not always available.
For all combinations of transmit power and payload length, we simulated the transmission of \SI{10000}{packets}, of which we selected a random sequence of \SI{5000}{packets} for each of the following analyses.
In a standalone simulation, we then implement the behavior of the different classes of MAC aggregation schemes and their selected parameterizations to extract which messages eventually become authenticated for a given binary loss trace.
Our measurements focus on the achieved goodput by the different \ac{MAC} aggregation schemes, where goodput is the ratio of received (and authenticated) payload bytes (\ie excluding header and authentication tag) to the number of transmitted bytes.
We initially focus on goodput as performance metrics as it directly measures how efficient the transmission channel is utilized, the improvement of which is the main goal of MAC aggregation.

\subsection{Influence of Channel Quality on Goodput}
\label{sec:synthetic:power}

\begin{figure}[t]
\includegraphics[width=\textwidth]{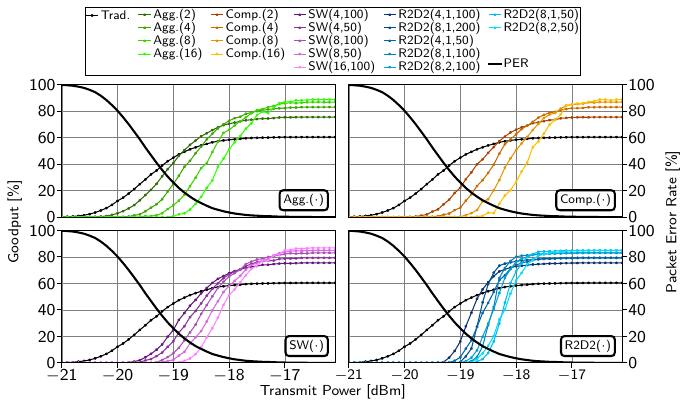}
\caption{
	Traditional MAC performs best with high packet error as all received data can be verified.
	For medium PERs, the aggregation of two tags with \agg and various \rd parameterizations are preferable, while the aggregation of more messages with the simpler \agg and \comp scheme are is desirable with low PERs.
} 
\label{fig:synthetic:power}
\end{figure}

For an initial understanding of the different MAC aggregation schemes, we fixed the payload length to \SI{48}{\byte} and gradually increased the transmission power, resulting in a slowly decreasing \ac{PER} from \SI{100}{\%} to \SI{0}{\%}.
Figure~\ref{fig:synthetic:power} shows our results.

We observe that all aggregation schemes exhibit the same general sigmoidal behavior: 
As the \ac{PER} decreases, the achieved goodput increases slowly before increasing quickly and then leveling off. 
This behavior can be explained by the behavior of the packet delivery ratio (\ie the opposite of the packet error rate), which also increases first slowly and then rapidly as the channel quality improves.
The interesting differences between the schemes and their parameterizations are thus defined by when and how goodput increases as the channel  improves.

For the different aggregation schemes, we see that the maximally achieved goodput correlates inversely with the number of aggregated tags (parameter $n$).
As a higher $n$ results in, on average, shorter tags, a better maximal goodput can be achieved due to less overhead.
Similarly, the transmit power where the goodput of the different schemes starts to take off also correlates with $n$.
The increasing likelihood can explain this observation that at least one of the tags in an aggregate cannot be computed as the set of aggregated messages becomes larger. 
Thus, the parameterizations with the higher bandwidth saving potential also require a better channel (\ie lower \ac{PER}) to be beneficial over more conservative parameterizations.
Consequently, traditional \acp{MAC} perform best with high \acp{PER} while exposing the overall worst goodput as \ac{PER} approaches \SI{0}{\percent}.

Comparing the performance of the different aggregation schemes, we observe that all schemes tend towards the same discrete goodput dictated by their average tag length.
However, the goodput provided by \agg, \comp, and \sw increases earlier but more slowly with increasing transmit power in contrast to \rd, which suddenly jumps up once the channel is good enough.
The behavior of \rd can be explained by ideally distributing the effects of packet loss to surrounding messages, such that if security levels for a few messages become good enough to consider the message authenticated, surrounding messages are close to the threshold as well.
Overall, for transmit powers up \SI{-18.9}{dBm} (PER=\SI{18.5}{\percent}), traditional \acp{MAC} perform best as they are not handicapped by the many lost packets.
Then, the aggregation of two messages with \agg is best until, between \SI{-18.3}{dBm} (PER=\SI{8.5}{\percent}) and \SI{-17.1}{dBm} (PER=\SI{0.4}{\percent}), there are different parameterizations of \rd that perform best.
As the \ac{PER} reduces further, the selected scheme becomes, however, less critical, and the differences for the same average tag length are marginal.
Here, simpler schemes with no overprovisioning, such as \agg and \comp, are usually preferable.
Consequently, it mostly depends on the channel quality, which aggregation scheme and parameterization achieve the best goodput.

\subsection{Influence of Payload Length on Goodput}
\label{sec:synthetic:length}

\begin{figure}[t]
	\includegraphics[width=\textwidth]{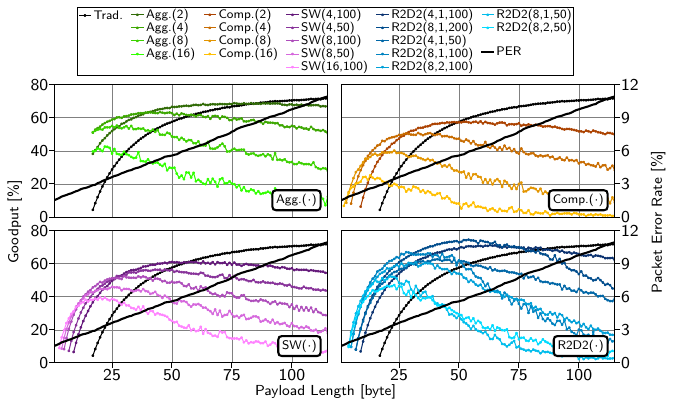}
	\caption{For larger payloads, the PER increases and the relative overhead of authentication tags decrease. Therefore, different schemes and parameterizations are optimal depending on payload lengths.} 
	\label{fig:synthetic:length}
\end{figure}

In Section~\ref{sec:synthetic:power}, we consider a fixed payload length and slowly increase the transmit power to improve the signal-to-noise ratio.
To better understand the behavior of the different MAC aggregation schemes, we now vary the payload length for a fixed transmit power of \SI{-18.3}{dBm}, where we have realistic \ac{PER} between 1.5 and \SI{10.9}{\percent} across the payload length range.
We show our results in Figure~\ref{fig:synthetic:length}.

With changing transmit power, we observe the same characteristics in the goodput curves of all aggregation schemes.
Goodput first quickly increases before slowly dropping after reaching a maximum.
This phenomenon can be explained by the overlapping effects of reduced relative overhead of authentication tags and growing numbers of unverifiable tags due to raised \acp{PER} with increased payload lengths.
Thus, selecting the best \ac{MAC} aggregation scheme depends on the underlying channel quality, packet lengths, and resulting variable \acp{PER}.

Furthermore, we can see that not all aggregation schemes can be employed for short payload lengths.
Traditional \ac{MAC} and \agg append 16-byte authentication tags (to a fraction of all) messages and thus require payload lengths of at least \SI{17}{\byte}.
The other aggregation schemes append a shortened tag to all messages, but the size of these tags also dictates how small messages can be.
Thus, if transmitted packets can only carry a few bytes of payload, such as the unreliable CAN bus protocol, which supports at most 8-byte payloads and has no header fields intended for integrity protection, the choice of available \ac{MAC} aggregation scheme shrinks.

Moreover, we observe different optimal payload lengths \wrt to goodput for the distinct schemes and parameterizations.
While using the maximal payload length of \SI{114}{\byte} yields the optimal goodput of \SI{71.7}{\%} for traditional MACs, the overall maximal goodput of \SI{74.4}{\%} is achieved by R2D2(8,1,200) with a payload length of \SI{54}{\byte}.
Hence, investigating the combined impact of packet lengths and MAC aggregation under varying conditions is essential to determine optimal network configurations in novel deployments.

\subsection{Optimal Packet Lengths for Authenticated Data}

\begin{figure}[t]
	\includegraphics[width=\textwidth]{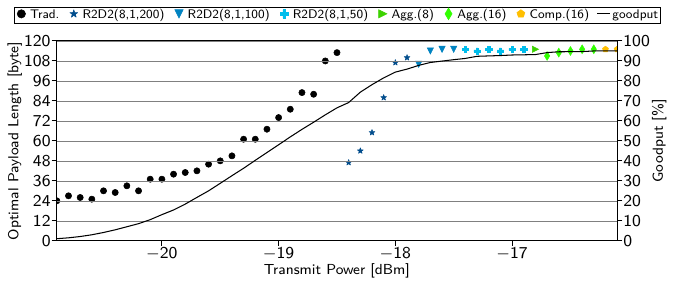}
	\caption{Different MAC (aggregation) schemes achieve a higher goodput as channel quality improves under optimal payload lengths. Unintuitively, changing a scheme can result in a reduced optimal payload length even if the channel improves.} 
	\label{fig:synthetic:optimal}
\end{figure}

Prior results indicate that considering the \ac{MAC} aggregation scheme is crucial when optimizing packet lengths for a given channel.
This search for optimal payload length gathered interest in the past~\cite{2003_sankarasubramaniam_energy,2008_vuran_cross,2016_akbas_joint,2016_kurt_packet} to make use of limited bandwidth availability or optimize the lifetimes of battery-powered devices.
As resource-constrained devices consume most of their power for wireless transmissions~\cite{2016_shaikh_radio_power}, optimizing goodput is essential for improving device lifetimes.
Assuming constant energy consumption for each transmitted bit at a given transmit power, the optimal combination of payload lengths and \ac{MAC} aggregation scheme also optimizes device lifetimes. 
These packet length optimizations, thus far, only looked at received data and not received \emph{and authenticated} data.
Assuming the imperative requirement of authenticated data, we search for the optimal payload lengths to optimize goodput across varying channel qualities, considering the different \ac{MAC} aggregation schemes.
Our results are shown in Figure~\ref{fig:synthetic:optimal}.

For low transmission powers, \ie low signal-to-noise ratio, we see that traditional \acp*{MAC},\ie no aggregation, perform best. 
This behavior can be explained by the initially high \ac{PER}, even for small messages, such that aggregated tags have a high risk of being composed of at least one message that did not arrive.
Here, the behavior observed in our setup matches related work~\cite{2003_sankarasubramaniam_energy,2008_vuran_cross,2016_akbas_joint,2016_kurt_packet} in that optimal payload lengths are initially short and then slowly increase as the transmit power is increased.

As the transmission channel improves, message aggregation starts to pay off since the benefits of shorter tags outweigh the risk of received data that cannot be authenticated.
Here, initially, between -18.4 and \SI{-17.2}{dBM}, \rd under various parameterizations performs best. 
However, the best MAC aggregation scheme does not only change with better channels; the optimal payload also decreases on each change before slowly increasing again.
Therefore, the optimal payload length for a transmit power of \SI{-18.5}{dBM} is \SI{113}{\byte} (with traditional \acp{MAC}), but for a slightly higher transmit power of \SI{-18.4}{dBM}, it drops down to \SI{47}{\byte} (for R2D2(8,1,100)). 
We can observe this same phenomenon for other changes between MAC aggregation schemes, and it is more or less pronounced depending on the header sizes, where the static overhead of larger packet headers dampens the drop in optimal packet sizes.

Overall, we can see that the average tag lengths of the optimal schemes shrink for higher transmission power.
Looking at the achieved goodput by the respective optimal scheme, we see a sigmoid curve that instead of leveling off at \SI{82.5}{\percent} if only using traditional \acp*{MAC}, the different MAC aggregation schemes boosts this achievable goodput to \SI{95.0}{\percent} as the \ac{PER} approaches 0.
However, it must also be understood that transmitting with optimal payload lengths is often not an option in practice.
Here, the (established) applications and protocols often dictate the payload lengths, \eg a sensor may only have a single reading that should be transmitted quickly and thus has no other data to fill into the payload.
Therefore, and because real wireless channels change over time, it is necessary to investigate MAC aggregation in real-world scenarios.

\section{MAC Aggregation in Real-World Scenarios}
\label{sec:real}

Thus far, we have analyzed MAC aggregation schemes in controlled synthetic environments.
While these analyses gave us insights into the behavior and nuances of the different schemes, they do not necessarily represent the entire story for realistic deployments.
Here, we often have predetermined payload lengths dictated by available data or protocol specifications.
Also, channel qualities vary dynamically over time, especially if some communication partners are mobile.
In the following subsection, we first introduce distinct real-world scenarios, which we subsequently use to evaluate and compare the performance of the \ac{MAC} aggregation schemes (\cf Section~\ref{sec:background:schemes}) under realistic conditions.

\subsection{Description of the Scenarios}
\label{sec:real:scenarios}

For our realistic measurements, we rely on network traces collected from real-world scenarios.
Each trace has constant payload lengths and transmission configurations, and we extract a binary loss trace of which transmitted packets have been correctly received or not.
This trace is then fed into our simulation to analyze the MAC aggregation schemes.
We summarize the scenarios in Table~\ref{tab:scenarios} and briefly introduce them in the following subsections.

\renewcommand{\tabcolsep}{0.8mm}
\begin{table}[t]
	\centering
	\begin{tabularx}{\textwidth}{lccccccc}
	\hline
	\textbf{\Tstrut Scenario} & \textbf{Duration} & \textbf{Protocol} &\textbf{Header} & \textbf{Data} & \textbf{\#pkts} & \textbf{PER} & \textbf{Src} \\
	\hline
	\Tstrut ICS & 8\,hours & IEEE\,802.15.4 &11\,B & 20\,B & 57648 & 4.79\% & \cite{2021_haenel_ge-params} \\
	\Tstrut Office & 22\,hours & BLE &10\,B& 32\,B & 79032 & 3.22\% & \cite{2021_haenel_ge-params} \\
	\Tstrut Smart City (sta.)& 131\,days & LoRaWAN &13\,B& 16\,B & 18790 & 1.97\% & \cite{2020_bhatia_loed} \\
	\Tstrut Smart City (mob.)& 250\,days & LoRaWAN &13\,B& 24\,B & 17415 & 7.09\% & \cite{2021_rademacher_path} \\
	\Tstrut Underwater & 327\,min & \footnotesize GUWMANET & 31\,bit & 16\,B & 334 & 16.46\% & \cite{2023_dol_salsa}\\
	\hline
	\end{tabularx}
	\vspace{1mm}
	\caption{ Limited bandwidth availability for integrity protection is a serious challenge across a wide range of lossy environments. }
	\label{tab:scenarios}
\end{table}

\subsubsection{Industrial Control System (ICS) Scenario.}
For the first scenario, we look at a measurement campaign of wireless communication in a \SI{3600}{\square\meter} production hall with nearly a billion transmitted packets~\cite{2021_haenel_ge-params}.
We select a single representative link from the various configurations using the IEEE~802.15.4 protocol with a payload length of \SI{20}{B}.
Our trace covers a total of \SI{8}{\hour} of traffic on a typical workday with an overall PER of \SI{4.79}{\percent}.
In this scenario, we observe primarily short bursts of packet loss with channel quality changing mostly over longer time windows (hours), while phases of high error rates (upwards of \SI{50}{\percent}) are possible for several minutes.

\subsubsection{Office Scenario.}
With the same measurement setup as for the ICS scenario, wireless links between nodes placed in various office rooms on a single floor have been measured~\cite{2021_haenel_ge-params}.
Here, we select a Bluetooth Low Energy (BLE) communication  link with \SI{32}{B} payloads over a \SI{22}{\hour} window during a workday.
We observe a relatively constant error distribution with short error bursts of a few packets each and an overall \ac{PER} of \SI{3.22}{\percent}.

\subsubsection{Smart City (Stationary) Scenario.}
Our first smart city scenario is based on the LoED dataset~\cite{2020_bhatia_loed}, where nine LoRaWAN gateways were placed in central London.
We focus on the 18790 packets transmitted by a single stationary sender and received by any of the gateways.
With an overall PER of \SI{1.97}{\percent}, we see primarily isolated packet loss due to long idle times between two transmissions, and the channel only experiences long-term changes in quality over several days, potentially due to altering weather conditions.

\subsubsection{Smart City (Mobile) Scenario.}
In this scenario, mobile LoRaWAN senders transmit to a total of nine stationary gateways for \SI{250}{days}.
Specifically, the sender was mounted to the top of a garbage truck driving through a \SI{200}{\square\kilo\meter} area in the city of Bonn~\cite{2021_rademacher_path}.
We observe burstier errors and overall channel qualities changing significantly over days and weeks.
The burstiness is likely due to the sender quickly entering and exiting the line of sight of a gateway, while the long-term changes changing presumably again relate to the weather conditions.

\subsubsection{Underwater Scenario.}
Finally, we consider acoustic underwater communication, with a trace of 334 16-byte messages being transmitted over \SI{327}{\minute} between two stationary nodes placed in the sea~\cite{2023_dol_salsa}.
The measurements were conducted during moderately rough weather conditions.
Despite an overall high \ac{PER} of \SI{16.46}{\percent}, most of these errors occurred during long bursts interspersed with periods of high packet delivery rates.

\subsection{Evaluating MAC Aggregation in Realistic Scenarios}
\label{sec:real:eval}

\begin{figure}[t]
	\includegraphics[width=\textwidth]{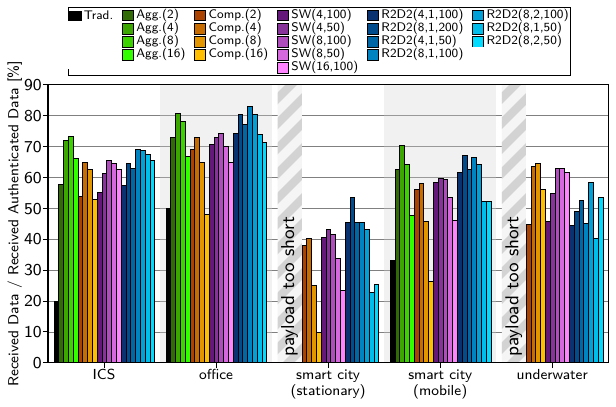}
	\caption{
		Different MAC aggregation schemes and parameterizations perform best, depending on the payload lengths, error burstiness, and overall PERs, such that the right scheme selection is non-trivial but crucial for the optimal use of constrained channels.
	} 
	\label{fig:real:goodput}
\end{figure}

We now analyze MAC aggregation schemes in the different realistic scenarios introduced in the previous section.
These scenarios are characterized by dynamic channels, differing communication protocols, and prespecified header and payload lengths.
For each scenario, we analyze the goodput (\ie the amount of received and authenticated data) in Figure~\ref{fig:real:goodput}.
We express the goodput as a percentage of the total amount of received payload data if no integrity protection was used.
For the urban (static) and underwater scenarios, traditional MACs and \agg cannot be included since the tags do not fit into the available payload.

We see different MAC aggregation schemes performing best in the synthetic measurements, depending on the investigated scenario.
Payload lengths influence the concrete parameterization but do not directly correlate with which scheme performs best under the relatively small variations observed across the different scenarios.
The  best-performing \ac{MAC} aggregation scheme  thus depends primarily on two factors: overall PER and burstiness.

For the industry and urban (mobile) scenario, where overall \ac{PER} is low and burstiness relatively high, \agg performs best.
During a burst, most packets in one set of aggregated messages are lost, while otherwise, most sets are received entirely and can be authenticated.
For the office and urban (static) scenarios, \rd performs best due to the high \ac{PER} and the short error bursts, where often only a single packet is lost.
However, higher \acp{PER} do not immediately mean that \rd performs best (until traditional \acp{MAC} are more favorable), as suggested by the synthetic scenarios.
The long burst, where no traffic passes, in combination with a relatively good delivery ratio otherwise mean that \sw and \comp perform best in the underwater scenario. 
Overall, the best \ac{MAC} aggregation scheme can achieve relative improvements of up to \SI{24.2}{\percent} better goodput compared to the second best scheme.

However, more important than selecting the scheme is using the correct parameterization.
If the wrong parameters are used for the best-performing \ac{MAC} aggregation scheme, performance can drop by an average between 14.0 and \SI{57.4}{\percent} in the worst case. 
With the \ac{PER} of the different scenarios ranging between 1.97 and \SI{16.46}{\percent}, we have parameterizations that result in average tag lengths of \SI{4}{\byte} performing best, which is more or less in line with the synthetic measurements from Section~\ref{sec:synthetic:power}.

Overall, we see that \ac{PER} and error burstiness play a significant role in finding the best scheme and parameterizations.
Due to relatively small variance in payload lengths across the scenarios, we, however, cannot confirm its low impact in selecting the best schemes.
Nevertheless, we know that the potential gains achieved through \ac{MAC} aggregation shrink with larger payloads.
Most importantly, we can conclude that adequate parameterization is more important than finding the best \ac{MAC} aggregation scheme. 
Ultimately, both optimizations have a non-negligible effect on the achievable goodput.

\section{Beyond Goodput as Evaluation Metric}
\label{sec:beyond}

Optimizing the goodput of \ac{MAC} aggregation is the main goal for most scenarios.
However, other effects must also be considered when choosing the \ac{MAC} aggregation scheme, such as verification delay, processing overhead, and susceptibility to jamming attacks.
In the following subsections, we compare the different \ac{MAC} aggregation schemes (\cf Sec.~\ref{sec:background:schemes}) \wrt these effects.

\subsection{Average Delay until Authentication}
\label{sec:beyond:delay}

\begin{figure}[t]
	\includegraphics[width=\textwidth]{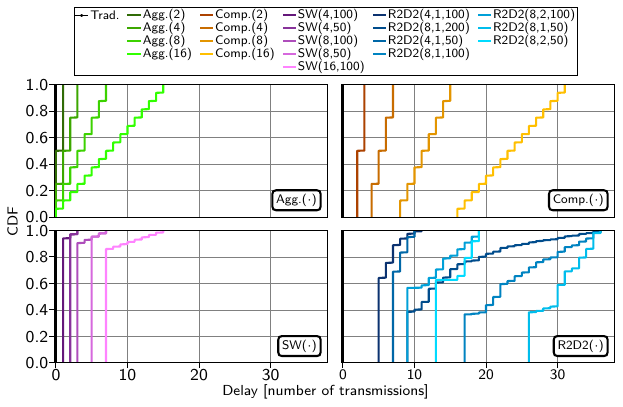}
	\caption{Verification delay is an inherent drawback of MAC aggregation. For scenarios where verification delay is critical, \sw does, however, provide highly consistent delays which control algorithms can thus anticipate.} 
	\label{fig:real:delay}
\end{figure}

First, we look at the authentication delay for the different \ac{MAC} aggregation schemes.
Traditional authentication tags can be verified immediately upon message reception, so no delay occurs due to waiting for additional data.
With \ac{MAC} aggregation, on the other hand, we need to wait until all messages depending on a specific tag have been received to verify it, which might introduce significant delays.
To analyze these effects, we plotted the delay from the measurements on all traces from Section~\ref{sec:real:scenarios} as a CDF in Figure~\ref{fig:real:delay}.

We see major differences in the behavior of the different aggregation schemes for these measurements.
\agg and \comp periodically verify a set of prior messages together, such that a range of different delays occur with the same frequency.
The concrete span of possible delay is then proportional to the parameter $n$ of how many tags are aggregated together.

\sw, on the other hand, verifies messages continuously with an almost constant delay.
This delay only varies if some messages get lost, which incurs a verification delay for surrounding messages.
This behavior is beneficial for applications requiring periodic messages with practically no jitter, \eg control algorithms in \acp{ICS} relying on a constant delay of the received information.
\rd shows similar behavior for about half of all the received messages, while the rest have increasing delays.
Again, the packet loss is responsible for higher delays, but since \rd distributed the effects of packet losses over multiple packets, more of them experience delayed verification.
Moreover, the magnitude of these delays correlated with the overprovisioning factor $o$, allowing late authentication for messages that could otherwise not be authenticated.

In summary, the average delay until authentication of the different authentication schemes strongly differs.
While \agg offers, on average, the lowest delays, \sw has the most constant delays.
On the other hand, \rd offers the best goodput for many scenarios with higher PER while messages have higher and more varying verification delays.
Selecting the best aggregation scheme according to this delay thus depends on which balance the concrete application scenario demands between the goodput reduction and the type of verification delay.

\subsection{Performance and Memory Overhead}
\label{sec:beyond:performance}

\begin{figure}[t]
	\includegraphics[width=\textwidth]{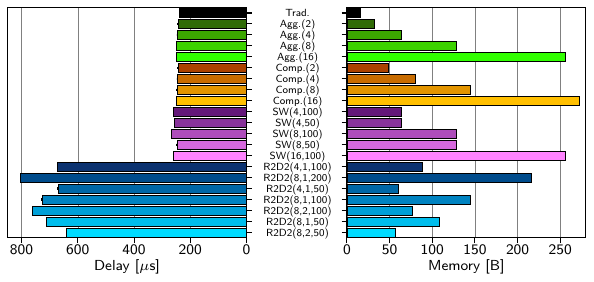}
	\caption{
		Only \rd introduces significant processing overhead over traditional MACs. Memory overhead, on the other hand, is mostly dependent on how many tags are aggregated together but so small that it should rarely be a decisive factor.
	} 
	\label{fig:real:performance}
\end{figure}

Many of the considered scenarios involve resource-constrained IoT devices where substantial additional processing and memory overhead from the \ac{MAC} aggregation scheme could significantly impact performance.
Hence, we measure and compare the processing delay and memory overhead for tag computation and buffering by the different schemes.
We conducted the analysis on the Arm Cortex M3 processor of a Zolertia RE-Mote board, a common choice to evaluate realistic resource-constrained hardware.
As a baseline, we capture the time to authenticate a single \SI{32}{\byte} message with hardware-accelerated HMAC-SHA256, which is the underlying MAC scheme used for the aggregation schemes as well.
We averaged the average processing times over \SI{16}{tag} generations (not all schemes do the same computations for each message) and repeated this measurement 30 times.
For the memory overhead, we measure the memory necessary to buffer tags before their aggregation, as all other memory overhead is implementation-dependent and is mostly optimized away by the compiler.
The results of both measurements are presented in Figure~\ref{fig:real:performance}.

Regarding processing times, we see only marginal overhead for all aggregation schemes except \rd.
There, we have a 168 to \SI{237}{\percent} increase in processing times compared to the baseline, where the differences across parameterizations are mostly insignificant.
This overhead stems from the bitwise processing of \rd, which requires a significant amount of XOR and bitshift operations.
This processing overhead is, however, mostly only impactful for applications that run on slower hardware and have tight latency requirements, especially considering that the sender \emph{and} receiver must conduct this additional processing.

We see a different picture across the MAC aggregation schemes for the memory overhead.
For \agg, \comp, and \sw, the needed memory depends on the message history.
More tags must be stored concurrently for shorter aggregated tags, resulting in higher memory overhead.
Here, the bitwise processing of \rd helps to partially process tags when new messages arise.
Consequently, the memory depends mainly on the overprovisioning factor and less on the number of aggregated tags.
The magnitude of the required additional memory for MAC aggregation schemes is, however, small enough that it should rarely influence the decision on which scheme should be deployed.

\subsection{Resilience to Adversarial Interference}
\label{sec:beyond:resilience}

In our final analysis, we compare the resilience of MAC aggregation schemes to selective jamming attacks.
Selective jamming refers to jamming specific messages to prevent their correct reception which enables stealthy and energy-saving attacks as dropped packets are hardly distinguishable from random packet loss~\cite{2011_wilhelm_reactive-jamming,2017_aras_jamming}.
In the context of MAC aggregation schemes, a sophisticated attacker can amplify the effects of a denial-of-service attack due to the employed MAC aggregation.
For example, for \texttt{Agg}$(16)$, it suffices to jam every 16\textsuperscript{th} packet to reduce the (authenticated) goodput of the channel to zero. 

\begin{figure}[t]
	\includegraphics[width=\textwidth]{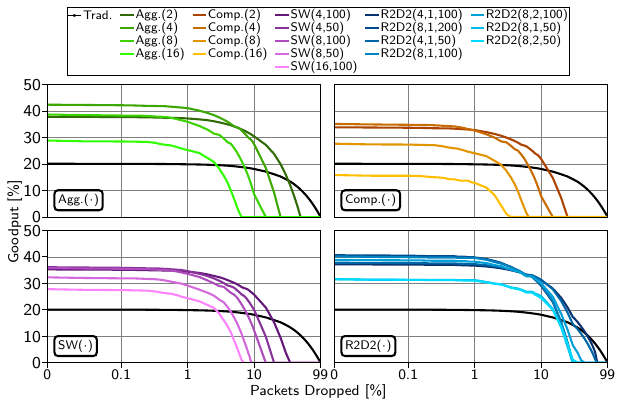}
	\caption{
		\rd shows significantly increased resilience to denial-of-service attacks through selective jamming, especially if an attacker jams less than \SI{10}{\percent} of messages to remain stealthy or conserve energy.
	} 
	\label{fig:beyond:resilience}
\end{figure}

For our measurements, we considered the trace from the urban (mobile) scenario introduced previously, as its payload is large enough for all schemes, and urban settings provide easy access to potential attackers.
For each aggregation scheme, we developed the optimal jamming attack strategy to minimize the goodput at the receiver.
In Figure~\ref{fig:beyond:resilience}, we show how the achieved goodput of the different aggregation schemes is impacted by increased attacking capabilities.

The x-axis represents the number of overall dropped packets in percent on a logarithmic scale.
For traditional authentication, we see, as expected, that the channel can still transmit authenticated data as long as not the entire channel is jammed.
In general, we note that shorter average tags are more susceptible to selective jamming attacks, as each tag requires many received messages to become verifiable.
Considering the shortest tags (n=16) for  \agg, \comp, and \sw, we see that dropping between 
27 and \SI{29}{\percent} targeted packet already suffices to prevent all data transmission over the channel.

The behavior of \rd requires, however, a separate analysis since one of the protocol's design goals is resilience against jamming attacks.
Therefore, the exact dependencies between tags and messages are kept secret, such that attackers can only design their strategy to inflict the most damage for the average dependency selection.
Furthermore, the design of \rd explicitly distributes the effects of packet losses (malicious or not) over many packets, thus cushioning the impact of selective jamming.
Hence, up to \SI{15}{\percent} of packets need to be dropped to reduce goodput by even \SI{20}{\percent}.
However, once a critical mass of packet loss occurs, such distribution no longer suffices for compensation, and the goodput quickly drops.

Overall, we can say that \rd is the most resilient scheme in the presence of a selective jammer.
Considering our entire analysis, no scheme is an outright winner, and each scheme has its benefits.
To summarize these findings, guide operators toward the right MAC aggregation scheme, and identify open research questions, we provide general recommendations in the following section.

\section{Guidelines on Employing MAC Aggregation}
\label{sec:guidelines}

In general, MAC aggregation shows promising potential to boost available bandwidth on lossy channels for various scenarios.
However, not every scenario benefits from MAC aggregation compared to traditional MACs.
More importantly, choosing the correct scheme and parameters is decisive in answering the questions of \emph{when} and \emph{how} to use MAC aggregation.
Therefore, in the following, we deepen this discussion towards providing general guidelines on employing MAC aggregation based on our empirical measurements.

\subsection{When to Use MAC Aggregation on Lossy Channels?}
From our analysis, it is evident that MAC aggregation reliably improves goodput for relatively high \acp{PER} of \SI{10}{\percent} or below. 
In cases where the PER is higher, it is often more beneficial to rely on traditional MACs or, at most, aggregate MACs for no more than two messages~(\ie setting the parameter $n$ to 2). 
However, for high PERs due to long error bursts where hardly any traffic arrives, MAC aggregation can still be beneficial~(\cf~Section~\ref{sec:real:eval}).

Furthermore, we investigated the relationship between payload lengths and the resulting benefits of MAC aggregation. 
For instance, in scenarios involving \SI{200}{\byte} payloads and minimalistic \SI{5}{\byte} headers, a MAC aggregation scheme aggregating 16 tags (\ie $n=16$) could still generate a \SI{7.3}{\percent} goodput improvement.
Consequently, we conclude that MAC aggregation, in general, offers the most substantial benefits for short payload lengths, up to a few hundred bytes, and moderate PERs of up to \SI{10}{\percent}.
As substantiated by the real-world scenarios~(\cf~Section~\ref{sec:real:scenarios}), this is precisely the kind of communication that occurs in many (industrial) \ac{IoT} scenarios, leading to the question of how to use MAC aggregation in such scenarios to gain the most benefit.

\subsection{How to Employ MAC Aggregation on Lossy Channels?}
In our evaluations of the goodput improvements that different MAC aggregation schemes and parameterizations can bring in real-world scenarios (\cf~Section~\ref{sec:real:eval}), 
we have seen that no aggregation scheme is a clear-cut winner (even when solely focusing on goodput as an evaluation metric).
Moreover, we have seen that the correct parameterization for a given scenario is crucial to achieving optimal performance.
These observations thus warrant a more nuanced discussion of when to use which MAC aggregation scheme and with which parameters.

Focusing solely on goodput, we see that generally \rd achieves the highest performance for \ac{PER} between 0.4 and \SI{8.5}{\percent}, especially when packet errors occur as short bursts.
For lower \acp{PER} and traffic with longer error bursts, the better performance and simplicity of \agg is often preferable.
If the periodic 16-byte tags for \agg are not supported by the application (\eg due to fixed message sizes), \comp is a good alternative to realize a constant tag size across all messages.
Considering the parameterizations, a high $n$ has the potential to realize better goodput, but only if the \ac{PER} is relatively low.
For the overprovisioning factor $o$ of \sw and \rd, $100$ is usually the best or least a decent choice.
\rd's $g$-factor is best set to 1 in those scenarios where \rd achieves the best goodput.
Overall, \sw rarely outperforms the other schemes if only considering goodput since it is not designed for lossy communication~\cite{2022_wagner_spmac}.
Nevertheless, it can still be a sensitive choice when also considering \eg verification delays.

One disadvantage of MAC aggregation compared to traditional MACs is the inherent verification delay which we investigated in Section~\ref{sec:beyond:delay}.
This delay occurs as most messages cannot be verified directly upon reception and thus need to be buffered or processed optimistically~\cite{2022_wagner_spmac}, \ie processed under the assumption of being genuine before full integrity verification.
This risk can be reduced by the two progressive schemes \sw and \rd, already providing some, yet reduced, security guarantees immediately upon message reception.
Furthermore, if an application requires complete message verification, \sw provides deterministic verification delays, beneficial for real-time control.

Concerning other potential dimensions for selecting the best MAC aggregation scheme for a given scenario,
memory overhead is so small that it should rarely be a decisive factor.
When interested in optimizing processing overhead, only \rd shows a clear disadvantage~(\cf~Section~\ref{sec:beyond:performance}) compared to the other aggregation schemes.
Finally, if resilience to denial-of-service attacks through selective jamming is essential, \rd shows clear advantages over the other schemes.
However, if another scheme must be used (\eg due to the excessive processing overhead of \rd), then lowering the parameter $n$ can reduce the effects of attacks at the cost of reduced goodput under normal operation.

\subsection{Selecting an MAC Aggregation Scheme}

\begin{figure}[t]
	\includegraphics[width=\textwidth]{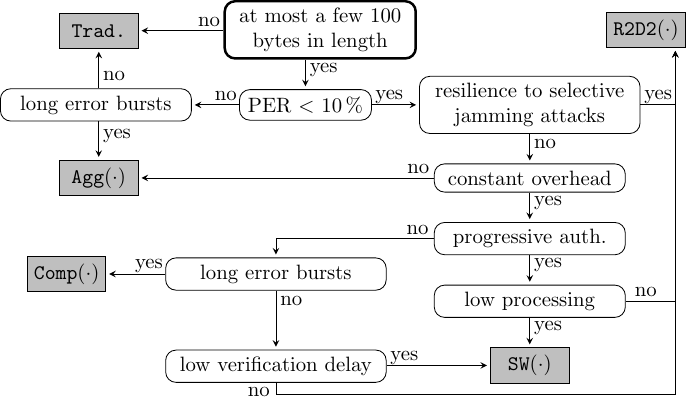}
	\caption{The optimal MAC aggregation scheme depends on many different characteristics. This decision diagram assists in this selection process.} 
	\label{fig:decision}
\end{figure}

We observe that often many dimensions must be considered to decide when and how to perform MAC aggregation.
To help operators in their decision process, we provide two forms of assistance.
First, we provide a decision diagram to select the right MAC aggregation scheme in Figure~\ref{fig:decision} based on basic network characteristics and feature demands.
Secondly, and for more detailed analysis, we provide an evaluation tool\footnote{\url{https://github.com/fkie-cad/mac-aggregation-analysis-tool}.} to aid further in this decision process.
Our evaluation tool takes as input the header and payload lengths as well as an example binary loss trace, \ie a series of 1s and 0s for received and dropped packets, respectively.
It provides a comparison of all MAC aggregation schemes and their parameterizations (as analyzed in this paper) for the given scenario.
In combination with these tools, our guidelines support operators in deciding when and how to employ MAC aggregation and help researchers to identify further opportunities to optimize existing MAC aggregation schemes.

\section{Conclusion}

MAC aggregation effectively saves valuable bandwidth in resource-constrained networks by shifting integrity protection from single to multiple packets.
However, as shown in this paper, the potential benefits of MAC aggregation strongly depend on the individual network scenario.
In particular, the effects of (bursty) packet losses, as experienced in wireless communication, severely impact the performance of MAC aggregation.
Therefore, we specifically address the research question of \emph{when} and \emph{how} to aggregate MACs by comparing existing aggregation schemes in synthetic and real-world scenarios.
Our empirical results indicate that, in general, MAC aggregation is particularly effective in scenarios with relatively reliable communication (\ie with PERs below \SI{10}{\percent}) and for short payload lengths (\ie below a few hundred bytes).
Most importantly, however, correctly parameterizing MAC aggregation is even more critical than choosing the right scheme. 
Moreover, other optimization metrics than goodput may limit the choice of applicable MAC aggregation schemes and thus need to be considered.
With our detailed guidelines and our public evaluation tool, we intend to support operators in deciding when and how to employ MAC aggregation for their applications and researchers to improve MAC aggregation further, ultimately strengthening security even under adverse networking conditions.

\section*{Acknowledgements}

We thank Thomas Hänel, Michael Rademacher, and Michael Goetz for providing access to their datasets.
We thank Jan Bauer, Michael Rademacher, and our anonymous reviewers for valuable feedback that improved this paper.
This work is partially funded by the project MUM2, partially funded by the German Federal Ministry of Economic Affairs and Climate Action~(BMWK) with contract number~03SX543B managed by the Project Management Jülich~(PTJ), and by the Deutsche Forschungsgemeinschaft (DFG, German Research Foundation) under Germany's Excellence Strategy -- EXC-2023 Internet of Production -- 390621612.
The authors are responsible for the contents of this work.

\bibliographystyle{splncs04}
\bibliography{paper}

\end{document}